# Adaptive hard and tough mechanical response in single-crystal B1 VN$_x$ ceramics via control of anion vacancies

A.B. Mei,[1] H. Kindlund,[2] E. Broitman,[3,4] L. Hultman,[4] I. Petrov,[4,5] J.E. Greene,[4,5,6]
D.G. Sangiovanni[4,7*]

[1]Department of Materials Science and Engineering, Cornell University, Ithaca, NY, 14853, USA

[2]Department of Materials Science and Engineering, University of California,
Los Angeles, CA 90095, USA

[3]SKF, Research & Technology Development, 3992AE Houten, Netherlands

[4]Department of Physics, Chemistry and Biology (IFM) Linköping University,
SE-581 83, Linköping, Sweden

[5]Department of Materials Science and the Materials Research Laboratory University of Illinois,
104 South Goodwin, Urbana, IL 61801, USA

[6]Department of Materials Science, National Taiwan University of Science and Technology,
Taipei 10607, Taiwan

[7]Interdisciplinary Centre for Advanced Materials Simulation (ICAMS),
Ruhr-Universität Bochum, D-44780 Bochum, Germany

High hardness and toughness are generally considered mutually exclusive properties for single-crystal ceramics. Combining experiments and *ab initio* molecular dynamics (AIMD) atomistic simulations at room temperature, we demonstrate that both the hardness and toughness of single-crystal NaCl-structure VN$_x$/MgO(001) thin films are simultaneously enhanced through the incorporation of anion vacancies. Nanoindentation results show that VN$_{0.8}$, here considered as representative understoichiometric VN$_x$ system, is ≈20% harder, as well as more resistant to fracture than stoichiometric VN samples. AIMD modeling of VN and VN$_{0.8}$ supercells subjected to [001] and [110] elongation reveal that the tensile strengths of the two materials are similar. Nevertheless, while the stoichiometric VN phase cleaves in a brittle manner at tensile yield points, the understoichiometric compound activates transformation-toughening mechanisms that dissipate accumulated stresses. AIMD simulations also show that VN$_{0.8}$ exhibits an initially greater resistance to both {110}⟨1$\bar{1}$0⟩ and {111}⟨1$\bar{1}$0⟩ shear deformation than VN. However, for progressively increasing shear strains, the VN$_{0.8}$ mechanical behavior gradually evolves from harder to more ductile than VN. The transition is mediated by anion vacancies, which facilitate {110}⟨1$\bar{1}$0⟩ and {111}⟨1$\bar{1}$0⟩ lattice slip by reducing activation shear stresses by as much as 35%. Electronic-structure analyses show that the two-regime hard/tough mechanical response of VN$_{0.8}$ primarily stems from its intrinsic ability to transfer *d* electrons between 2$_{nd}$-neighbor and 4$_{th}$-neighbor (i.e., across vacancy sites) V–V metallic states. Our work offers a route for electronic-structure design of hard materials in which a plastic mechanical response is triggered with loading.

**\*Corresponding author:** davide.sangiovanni@liu.se

# 1. Introduction

Brittle fracture in ceramics is primarily caused by the limited ability of these materials to dissipate mechanical stresses ahead of a growing crack [1, 2]. Accordingly, a strategy adopted to mitigate brittleness involves enhancing hardness while simultaneously hindering or deflecting crack propagation. This is commonly achieved through grain-boundary [3] and nanostructure engineering [4, 5]. However, recent experimental results [6-9] demonstrated that single-crystal NaCl-structure (B1) pseudobinary $V_{0.5}Mo_{0.5}N$ transition-metal (TM) nitride ceramics are intrinsically *both* hard (~20 GPa) and ductile. This surprising finding confutes the commonly accepted assumption that excellent ductility and high strength (well-correlated to hardness in solids [10, 11]) are mutually-exclusive material properties [12, 13]. Experiments have also proven that single-crystal B1 $V_{0.5}Mo_{0.5}N_x$ solid solutions become much harder (from 17 to 26 GPa) when the concentration of anion vacancies increases up to 45% [7].

Previous experimental results show that the hardness (H) of B1 Group-VB (i.e. V, Nb, and Ta) nitrides and carbides is enhanced by anion vacancies [14-17], consistent with H vs. x trends in $V_{0.5}Mo_{0.5}N_x$ [7-9]. Nonetheless, although understoichiometric $V_{0.5}Mo_{0.5}N_x$ is considerably less prone to crack than single-crystal B1 TiN and B1 VN, it is more susceptible to fracturing than stoichiometric $V_{0.5}Mo_{0.5}N$ [7]. This raises the question of whether hardness and ductility can be enhanced at the same time by controlling the metal/non-metal compositional ratio in refractory carbonitrides.

In this work, we combine experiments and *ab initio* molecular dynamics (AIMD) atomistic simulations to demonstrate that control of the N/V stoichiometry in $VN_x$ allows simultaneously enhancing hardness, ductility, and toughness. The experiments are conducted on high-quality single-crystal B1 $VN_x$ deposited epitaxially on MgO(001) with nitrogen-to-vanadium ratios x spanning 0.8 – 1.0 [18-20]. Nanoindentation testing show that the unusual combination of remarkable mechanical properties in understoichiometric $VN_{0.8}$ – in preliminary tests, $VN_{0.8}$ exhibits the best performance among all investigated $VN_x$ systems – unambiguously originates from anion vacancies. AIMD simulations of VN and VN$_{0.8}$ subjected to tensile deformation (up to fracture) and uniform shearing (up to lattice slip) at 300 K are used to unravel key atomistic and electronic mechanisms responsible for the superior mechanical behavior induced by vacancies. The results provide unprecedented insights for simultaneously enhancing hardness, ductility, and toughness in single-crystal TM carbonitride ceramics simply by controlling the lattice stoichiometry.

## 2. Methods

**2.1. Experimental.** The B1 structure of VN$_x$ compounds has a large single-phase field that can accommodate compositional variations $0.7 < x \leq 1$, possible through incorporation of anion vacancies. VN$_x$/MgO(001) thin films with x spanning from 0.8 to 1.0 are grown to a thickness of 300 nm under 20 mTorr mixed N$_2$/Ar atmospheres in an ultra-high-vacuum magnetically-unbalanced reactive magnetron sputter-deposition system. Anion sublattice occupancy is controlled by varying the N$_2$ gas fraction between 0.1 and 1.0 and the growth temperature between 430 and 540 °C [19]. Rutherford backscattering spectroscopy is employed to determined nitrogen-to-vanadium ratios to an accuracy of ±0.05. MgO is selected as substrate material because it is isostructural with VN and exhibits a sufficient film/substrate lattice mismatch (0.9%) to enable film relaxation, a necessity for the investigation of intrinsic properties.

VN$_x$(001) hardness and Young's moduli values as a function of x are quantified [21, 22] via nanoindentation experiments performed on 300-nm-thick films in a Hysitron TI950 Triboindenter using a sharp Berkovich 142.3° diamond probe (tip radius ~150 nm) calibrated to an epitaxial TiN/MgO(001) sample. Nine indentations, arranged in a 3 × 3 pattern with indents separated by 10 µm, are made in each sample with a maximum tip penetration limited to 10% of the film thickness, which was verified to avoid spurious substrate effects on measured hardness values. The fracture toughness of VN$_x$ samples is assessed via nanoindentation using a cube-corner diamond probe, which is sharper than a Berkovich tip and thus results in higher contact stresses [23]. We performed two different tests: indentations at constant depths of 400 nm (which exceed the film thickness by 100 nm) and indentations as a function of penetration depth between 25 and 400 nm, at steps of 25 nm. At least 9 cube-corner indentations with constant depths of 400 nm were performed on each sample. The massive deformation induced in the films allows us to demonstrate the completely different mechanical behavior of stoichiometric VN (brittle) vs. understoichiometric $VN_{0.8}$ (supertough).

That compositional variations in understoichiometric $VN_x$ are accommodated through anion vacancies is concluded based on the evolution of $VN_x$ relaxed lattice parameters $a_0$ and mass densities $\rho$ as a function of x. $a_0(x)$ values determined from x-ray diffraction (XRD) high-resolution reciprocal-space-maps decrease from 0.4134 nm (x = 1.0) to 0.4087 nm (x = 0.8) (details given in Ref. [19]) – a trend which is accurately described with first-principles density-functional theory (DFT) results obtained as a function of nitrogen vacancy concentrations [19]. $VN_x$/MgO(001) mass densities, obtained from x-ray reflectivity scans as well as Rutherford backscattering spectroscopy combined with film thickness measurements [20], decrease from 6.06 (x = 1.0) to 5.98 g·cm$^{-3}$ (x = 0.76) [20]. $VN_{0.8}$ is expected to have a mass density of ≈6.0 g·cm$^{-3}$, in close agreement with the experimentally determined value; V interstitials and $V_N$ antisite substitutions would generate larger $\rho$ values of 7.4 and 6.6 g·cm$^{-3}$, respectively.

**2.2. Modeling.** Tensile and shear mechanical testing of $VN_x$ systems are modelled using density-functional *ab initio* molecular dynamics [24] simulations at room temperature.

Tensile-testing simulations enable the evaluation of ideal tensile strength $\gamma_T$ and toughness $U_T$ for different strain directions. Deformation is applied orthogonal to (001) and (110) crystal surfaces, for which the energy of formation is lowest in B1-structure ceramics [25]. In AIMD modeling, $\gamma_T$ corresponds to the maximum $\sigma_{zz}$ stress reached during elongation, whereas $U_T$ is calculated as the area that underlies the full stress/strain curve. Shear-deformation modeling is used to determine the resistance to change of shape of the materials, which is well-correlated with their hardness [26-28]. In addition, these simulations shed light on the mechanisms that govern the activity of {110}⟨1̄10⟩ and {111}⟨1̄10⟩ slip systems, typically operative in B1-structure nitrides and carbides at room and/or elevated temperatures [29-31].

AIMD simulations are carried out with VASP [32] implemented with the projector augmented-wave method [33]. The Perdew-Burke-Ernzerhof (PBE) electronic exchange and correlation approximation [34], Γ-point sampling of the Brillouin zone, and planewave cutoff energies of 300 eV are used in all simulations. The ionic equations of motion are integrated at timesteps of 1 fs, using $10^{-5}$ eV/supercell convergence criteria for the total energy during both system equilibration and mechanical testing. Prior to modeling tensile and shear deformation, the supercell equilibrium structural parameters are determined via NPT simulations at 300 K using the Parrinello-Rahman barostat [35] and the Langevin thermostat. Thus, NVT sampling of the configurational space is used to equilibrate the unstrained structures for 5 additional ps using the Nose-Hoover thermostat. In these simulations, it is ensured that the time-averaged $|\sigma_{xx}|$, $|\sigma_{yy}|$, and $|\sigma_{zz}|$ stress components are ≲ 0.5 GPa.

Our model supercells are oriented in a convenient manner for investigating the response of VN and VN$_{0.8}$ to tensile and shear deformation along different directions. Simulation boxes with [001] and [110] vertical (z) orientation are used to probe the dynamics of [001] and [110] tensile elongation, respectively. The supercells employed for {110}⟨1̄10⟩ and {111}⟨1̄10⟩ shear deformation (up to the activation of lattice slip) have the lateral (x) axis parallel to a Burgers vector direction ⟨1̄10⟩ and the vertical (z) axis orthogonal to a slip plane, i.e. {110} or {111}. VN and VN$_{0.8}$ supercells with vertical [h k l] orientation are denoted below as VN$_x$(h k l).

Tensile testing is modeled for VN$_x$(001) and VN$_x$(110) supercells with 768 lattice positions (24 layers orthogonal to the strain direction; 32 sites per layer), following the procedure of Ref. [36, 37]. Briefly, at each $\delta_{zz}$ strain step (1% of the supercell vertical size), the systems are first coupled to an isokinetic, and then to a Nosé-Hoover, thermostat. The thermostat temperature is set to 300 K, for a total equilibration time of 3 ps. The structures are progressively elongated until they reach their fracture points. VN and VN$_{0.8}$ simulation boxes used for shear deformation are formed of 576 lattice sites (24 layers parallel to the slip plane; 24 atomic positions per layer). The cells are equilibrated and sequentially deformed (shear strain steps $\delta_{xz}=0.02$), with procedure analogous to that used for tensile testing. In all VN$_{0.8}$ supercells, the vacancy distribution on the anion sublattice yields negligible degrees of short-range ordering, which is consistent with experimental analysis [19].

Average tensile $\sigma_{zz}$ and shear $\sigma_{xz}$ stresses are determined as a function of strain from 500 equilibrated AIMD configurations. We note that, while in previous *static* first-principles stress/strain calculations thermal disorder was mimicked by introducing arbitrary atomic displacements [27, 38], our simulations explicitly include temperature effects, thus providing a realistic description of phase transformations and non-linear elastic effects. The approach is particularly important for modeling crystal structures that are dynamically-stabilized by lattice vibrations, including B1 VN [39].

The simulation results are visualized using the VMD software [40]. Electron-transfer and energy-resolved electron-density maps are calculated for atomic configurations directly extracted from AIMD simulations at 300 K. Electron transfer is obtained by subtracting non-interacting atomic charge densities from the self-consistent electron density of the entire crystal. The electron densities in next-nearest-neighbor d-t$_{2g}$ and 4$_{th}$-neighbor (across-vacancy) d-e$_g$ metallic V states are visualized by resolving the total charge density into the energy intervals [-2 – 0 eV] and [-3 – -2 eV], respectively, where 0 corresponds to the Fermi level ($E_F$).

## 3. Results and discussion

**3.1. Characterization and mechanical properties of VN$_x$ films. Fig. 1a** shows a typical XRD ω-2θ scan acquired using Cu K$_{\alpha 1}$ radiation (wavelength $\lambda$ = 0.154056 nm) from an understoichiometric VN$_{0.8}$/MgO(001) layer. Only four peaks are observed: MgO 002 and 004 at 2θ = 42.8° and 94.0° and VN$_{0.8}$ 002 and 004 at 2θ = 43.9° and 96.8°. The absence of additional reflections, together with glancing angle and pole-figure scans (not shown), establish that VN$_x$(001) layers are NaCl-structure single-crystal films epitaxially oriented cube-on-cube relative to their MgO(001) substrate, $(001)_{VN_x} \parallel (001)_{MgO}$ and $[100]_{VN_x} \parallel [100]_{MgO}$. These findings are confirmed by high-resolution cross-sectional transmission electron microscopy (HR-XTEM) results, including the micrograph shown in **Fig. 1b**.

In our single-crystal B1 $VN_x(001)$ samples, understoichiometry is accommodated by vacancies on the anion sublattice [19]. Nanoindentation testing demonstrates that $VN_{0.8}$ is approximately 20% harder than VN ($H_{VN_{0.8}} = 17.1\pm0.8$ vs. $H_{VN} = 14.0\pm0.8$ GPa). This finding is consistent with previous experimental results, which have shown that the hardness of B1 Group-VB nitrides, including $VN_x$, increases with a N content that decreases from 100 to ≈80% [14-17]. The increased hardness of $VN_{0.8}$ is attributed to reduced dislocation mobility, which is also observed indirectly through the incomplete strain relaxation (92%) of $VN_x$ layers with $x \geq 0.05$ vs. $0 \leq x \leq 0.05$ (97% nitrogen) [19]. Thus, nitrogen vacancies obstruct dislocation glide (at least for relatively low loads; see the results of AIMD shear deformation below) resulting in higher hardness. As discussed in conjunction with AIMD-calculated elastic constants and moduli (see **Sec. 3.2** below), $VN_{0.8}$ exhibits *greater* shear elastic stiffness, but *lower* Young's moduli than stoichiometric VN.

Mechanical testing of single-crystal films demonstrates that $VN_{0.8}$ is not only harder, but also tougher than the stoichiometric compound. Vacancy-induced toughening in understoichiometric $VN_{0.8}$ layers is experimentally demonstrated by analyzing nanoindentation load-displacement curves (**Fig. 2**) and assessing indentation-induced fracturing using scanning-electron microscopy (SEM) in **Fig. 3**. Load-displacement curves L(d) are generated by indenting $VN_x$/MgO(001) layers with x = 1.0 and 0.8 using a Berkovich diamond tip. For shallow indents (d < 10 nm), L increases as $d^{3/2}$, indicating that deformation in this region is purely elastic [41]. Indeed, SEM imaging shows that retracting the tip while operating in this regime leaves no impression on the sample surface. For deeper indents (d>10 nm), a striking difference is observed by varying the vacancy concentration. For stoichiometric VN, plastic deformation is indicated by pop-ins – stochastically occurring displacement discontinuities [41, 42] – which reflect catastrophic mechanical failures on mesoscopic scales, including crack nucleation [43-45] and dislocation avalanches [46]. In contrast, nanoindentation load-displacement curves for understoichiometric $VN_{0.8}$ layers never exhibit pop-ins. Instead, L(d) gradually evolves into the elastoplastic regime through a series of small incremental displacements, suggesting that indentation stresses are continuously curbed below pop-in thresholds via the constant emission of dislocations.

**Fig. 3a-b** show representative SEM images in plan-view of 400-nm-deep indents (extending 100 nm into the MgO(001) substrate). Cube-corner nanoindentation results at constant penetrations of 400 nm demonstrate that $VN_{0.8}$ is considerably more resistant to fracture than VN. Indents into stoichiometric VN layers, **Fig. 3a**, exhibit 0.5-µm-long cracks propagating along [–110] directions. In contrast, understoichiometric $VN_{0.8}$ films *never* fracture. The absence of cracks in $VN_{0.8}$ layers is consistent with the continuous nanoindentation load-displacement curves observed for these samples and demonstrates their enhanced toughness. The integrity of the samples and absence of delamination at the film/substrate interface is assessed via cross-sectional SEM. A vertical section through an indentation site in stoichiometric VN/MgO(001), obtained with a focused $Ga_+$ ion beam in an FEI DualBeam 235 instrument, is presented in **Fig. 3c**. The indentation is triangular with a 0.5-µm-wide base. The tip of the indent lies 360 nm below the original film surface, indicating that 40 nm of the 400-nm-deep indent recovered elastically. The deformed VN/MgO(001) interface is 100 nm below the original interface. Despite the depth of the deformation, conformal contact is maintained along the entire film/substrate interface, indicating robust layer adhesion. Material pile-up around the indent occurs within 0.5 µm of indentation edges and rises 40 nm above the original surface.

Scanning probe microscopy (SPM) height isointensity contour maps show that plastic flow in indented VN samples occurs in the shape of a cross (**Fig. 3d**). Cube-corner indentation produces pronounced material pile-up (red areas in **Fig. 3d**) along ⟨110⟩ directions and less pile-up along ⟨100⟩. Mounds near the faces of the indentation triangle, where stress is concentrated, have heights (≈50 nm) which are more than twice those (≈20 nm) of hillocks near the vertices of the indentation triangle (**Fig. 3d**). Azimuthal rotations of the indentor with respect to the orientations shown in **Fig. 3**, changes relative mound heights, but has negligible effects on the pattern shape, in agreement with finite element models [47]. Pile-up patterns on $VN_{0.8}(001)$ have similar shapes but exhibit smaller mounds, consistent with their higher hardness. The higher degree of material pile-up observed for VN is also reflected in a lower elastic recovery of these films. Using all the collected load-displacement data – 12 load/unload curves for each sample (not shown) – we estimate an elastic recovery of 42±1% for VN and 55±1% for $VN_{0.8}$. We note, however, that the value of elastic recovery calculated for VN is reduced by the fracturing (large pop-ins) of the films during loading (**Fig. 2a**).

The analysis of the residual stresses present in $VN_x$ films is further in favor of our vacancy-induced toughening argument. At room temperature, the stoichiometric VN films have an in-plane residual compressive stress of -0.77 GPa. Although these compressive stresses should contribute to close the cracks upon unloading, the cracks remain visible after cube-corner nanoindentation of VN films. Conversely, the understoichiometric films have an in-plane residual tensile stress of +0.92 GPa. The fact that $VN_{0.8}$ films never crack despite of the presence of residual tensile stresses (which should favor crack formation) is a further proof of the excellent toughness of $VN_{0.8}$.

**3.2. AIMD evaluation of $VN_x$ mechanical properties at room temperature.** AIMD modeling is used to evaluate intrinsic mechanical properties of B1 $VN_x$ single crystals, and identify atomistic and electronic mechanisms responsible for the enhanced toughness and hardness of $VN_{0.8}$. Despite the fact that nanoindentation mainly produces compressive stresses in $VN_x$ samples, the regions of the films that surround the nanoindenter edges are subjected to severe tensile stresses. Indeed, brittle nitrides crack in proximity of nanoindender edges and corners (see **Fig. 3a** for VN and figure 3(c,d) in Ref. [6] for $VN_{0.89}$ and $TiN_{0.96}$). Accordingly, AIMD tensile testing is useful to assess the inherent materials' resistance to brittle fracture.

The stress/strain curves determined for supercells subjected to [001] elongation (**Fig. 4a**) show that VN(001) and VN$_{0.8}$(001) have ideal strengths (maximum stresses withstood during deformation) $\gamma_{TVN(001)}$=36 GPa and $\gamma_{TVN_{0.8}(001)}$=32 GPa. Strain along the [110] axis yields $\gamma_{TVN(110)}$=46 GPa and $\gamma_{TVN_{0.8}(110)}$=42 GPa, **Fig. 4b**. Hence, AIMD results show that the ideal tensile strength $\gamma_T$ of VN$_{0.8}$ is within ≈10% that of stoichiometric VN. However, VN$_{0.8}$ exhibits a much higher toughness and resistance to fracture than VN, with the most remarkable differences observed for [001]-strained compounds (**Fig. 4a**).

At an elongation of 14%, VN(001) cleaves on the (001) plane by sudden and essentially simultaneous breakage of V–N bonds parallel to the loading direction (**Figs. 5a and 4a**). In contrast, VN$_{0.8}$(001) resists fracture up to an elongation of 30% (**Figs. 5b and 4a**). The crack develops between (001) lattice layers characterized by a relatively *low* concentration of anion vacancies. In addition, VN$_{0.8}$ displays a total tensile toughness [48] which is approximately twice that of VN. The simulations reveal that the considerably enhanced resistance to fracture of VN$_{0.8}$ is due to a transformation toughening mechanism, characterized by buckling of (001) atomic planes, activated by extreme tensile stress (**Fig. 5b**). Analogous to the results obtained for (001) supercells, VN$_{0.8}$(110) also displays larger toughness than VN(110) (**Fig. 4b**) due to changes in bonding geometry activated at high tensile loading (AIMD snapshots not shown). The progressive transformation in the bonding network allows VN$_{0.8}$ to prevent stress build up, thereby hindering crack nucleation and actively toughening the material. The mechanism is similar to the one observed during AIMD mechanical testing [36] of hard and ductile B1 V$_{0.5}$Mo$_{0.5}$N$_x$ alloys [7]. We note that, within the supercell model considered here, the structural transformations identified in VN$_{0.8}$ compounds are fully reversible upon relaxation; after cleavage on (001) and (110) planes, VN$_{0.8}$ recovers the B1 structure.

As demonstrated by our nanoindentation tests, single-crystal B1 VN$_{0.8}$ samples (H$_{VN_{0.8}}$=17.1±0.8 GPa) are harder than stoichiometric films (H$_{VN}$=14.0±0.8 GPa). Detailed understanding of vacancy-induced hardening in VN$_x$ single crystals would require the challenging determination of representative dislocation-core structures – with their densities and nucleation mechanisms – as well as the kinetics of dislocation motion and dislocation/dislocation interactions [49-53]. Nevertheless, *ab initio* and experimental studies show that the trends in hardness of B1 TM nitrides and carbides are well correlated with their trends in shear elastic moduli [26, 28, 54, 55]. The fact that the hardness benefits from increased shear resistance is consistent with the results our experiments and AIMD simulations (see below).

**Fig. 6** illustrates the shear stress $\sigma_{xz}$ calculated for VN and VN$_{0.8}$ crystals subjected to [1$\bar{1}$0] shearing of (110) and (111) crystallographic planes. The peculiar upward bending of $\sigma_{xz}$ vs. strain $\delta_{xz}$ curves observed for small ($\delta_{xz} < 0.1$) shear deformations of VN(110) and VN(111) (**Fig. 6a-b**) is due to the large third-order elasticity of VN [56] (as elaborated in the following paragraphs). Conversely, the dependences of shear stresses $\sigma_{xz}$ in understoichiometric VN$_{0.8}$(110) and VN$_{0.8}$(111) are quasilinear for $\delta_{xz} < 0.1$ (**Fig. 6a-b**). The calculated $\sigma_{xz}$ values demonstrate that, within the elastic-response regime, VN$_{0.8}$ is more resistant to both {110}⟨1$\bar{1}$0⟩ and {111}⟨1$\bar{1}$0⟩ shear deformation than VN (note in **Fig. 6a-b** that $\sigma_{xzVN_{0.8}} > \sigma_{xzVN}$ for $\delta_{xz} < 0.09$). Hence, VN$_{0.8}$ exhibits an *initially harder* response to change of shape than VN.

To elucidate the differences in measured hardness values, it is worth to closely analyze the elastic responses to deformation of B1 VN$_x$ crystals. **Table 2** summarizes the results of the elastic constants and moduli evaluated by AIMD mechanical testing. The stiffnesses $C_{11}$, $G_{110}$, and $G_{111u}$ – calculated from the slopes (linear regression for ≤2% deformation) of stress/strain curves presented in **Figs. 4a, 6a, and 6b**, respectively – quantify the *actual* elastic resistances of B1 VN$_x$ crystals to ⟨001⟩ uniaxial tension, {110}⟨1$\bar{1}$0⟩ and {111}⟨1$\bar{1}$0⟩ shearing at room temperature. **Table 2** also lists *effective* average elastic constants $\widetilde{C}_{11}$, $\widetilde{C}_{12}$, and $\widetilde{C}_{44}$ obtained by minimizing the least-square differences between the stress components extracted during AIMD simulations and those predicted according to the linear elastic theory approximation [57]. The differences between *actual* and *effective* elastic properties are indicative of the magnitude of nonlinear elastic effects. Indeed, an ideal linear elastic response would yield essentially the same elastic constant values irrespectively of the choice of the strain tensor used for the calculation [56]. That also implies validity of the equivalences $\widetilde{C}_{ij} \equiv C_{ij}$ and that *actual* shear stiffnesses are accurately evaluated using the relations $G_{110} \equiv \widetilde{G}_{110} \equiv (C_{11}-C_{12})/2$ and $G_{111u} \equiv \widetilde{G}_{111u} \equiv (C_{11}-C_{12}+C_{44})/3$ [58, 59].

The *effective* average elastic properties $\widetilde{C}_{ij}$ determined for VN and $VN_{0.8}$ crystals are consistent with the range of values reported by acoustic waves measurements and DFT calculations at 0 K [18, 85-94] (see **Table 2**, [60, 61]). All $\widetilde{G}_{hkl}$ values are calculated as linear combinations of *effective* elastic constants (see end of previous paragraph). However, due to strong nonlinear elastic effects [56], the *actual* shear stiffnesses $G_{110}^{VN}$ (116 GPa) and $G_{111u}^{VN}$ (94 GPa) obtained via AIMD shear deformations of stoichiometric VN are ≈35%–50% *smaller* than $\widetilde{G}_{110}^{VN}$ (172 GPa) and $\widetilde{G}_{111u}^{VN}$ (193 GPa). Oppositely, the *actual* shear resistances of the understoichiometric compound ($G_{110}^{VN_{0.8}}$=205 GPa and $G_{111u}^{VN_{0.8}}$=190 GPa) are 20–25% larger than $\widetilde{G}_{hkl}^{VN_{0.8}}$ values (**Table 2**).

Nonlinear elastic effects are characteristic for anharmonic materials, as B1 VN [62], which are dynamically stabilized by lattice vibrations at finite temperatures [39, 63]. In the case of $\{110\}\langle1\overline{1}0\rangle$ shear deformation, the initial upward curvature of the stress/strain dependence of VN (**Fig. 6a**) can be rationalized as a deformation-induced softening of transversal acoustic phonon modes. The $\{110\}\langle1\overline{1}0\rangle$ shear deformation effectively corresponds to a pure tetragonal distortion of the cubic lattice (see illustration in figure 4 of [27]). The distortion is energetically facilitated by the B1→tetragonal VN transformation, which is a martensitic transition caused by phonon instabilities of the cubic lattice below 250 K (see schematic representation in figure 5 of [64], and results of electron and x-ray diffraction in figures 6 and 7 of [39]). On the contrary, experimental results indicate that an anion vacancy concentration of 3% thermodynamically stabilizes the B1 structure of $VN_x$ at cryogenic temperatures [63].

We propose that the greater shear elastic stiffnesses of $VN_{0.8}$ explain its experimentally-measured higher hardness in relation to VN. In addition to being harder than VN, the understoichiometric compound exhibits the remarkable capability to adapt its mechanical response to the loading condition. As the shear strain increases beyond ≈9%, $VN_{0.8}$ becomes progressively more compliant to change of shape. This facilitates plastic deformation in $VN_{0.8}$, as evidenced by the smaller ideal shear strengths γs and strains δs required to activate $\{110\}\langle1\bar{1}0\rangle$ and $\{111\}\langle1\bar{1}0\rangle$ slip in $VN_{0.8}$ vs. VN (**Fig. 6 and Table 1**).

Although $VN_{0.8}$ displays *higher* hardness and shear elastic stiffnesses than VN, both AIMD results and nanoindentation measurements show that $VN_{0.8}$ has a *smaller* Young's modulus than VN. The Young's moduli determined using the effective elastic constant values in **Table 2** in combination with the Voigt-Reuss-Hill average are $E_{VN}$ = 394±51 GPa and $E_{VN_{0.8}}$ = 342±55 GPa. The nanoindentation elastic moduli of VN and $VN_{0.8}$ – evaluated according to expression (2) in Ref. [21], together with $VN_x$ Poisson's ratios calculated using the AIMD elastic properties in **Table 2** – are $E_{VN}$ = 333±8 GPa and $E_{VN_{0.8}}$ = 316±11 GPa. The relatively smaller Young's moduli of $VN_{0.8}$ is also reflected in the lower steepness of the initial slope of load/displacement curves in **Fig. 2**. We note that a higher hardness in combination to a lower Young's moduli, which is the case of $VN_{0.8}$ vs. VN, has been proposed as empirical indicator for improved toughness in ceramic coatings.

The results of AIMD simulations allow us understanding the excellent resistance to fracture of $VN_{0.8}$ in comparison to the brittleness of VN (**Fig. 3**). The propagation of atomic-scale cracks in, and fracture toughness of, solid crystals depends in a non-trivial manner on the loading condition, crack and lattice geometries, and distribution of stresses within the material [65, 66]. Generally speaking, however, crack propagation can be prevented (crack blunting) if the stress concentrated around the crack tip is rapidly dissipated by plastic flow [67]. Crack blunting may occur if the emission of dislocations at an angle to the fracture plane is sufficiently fast, that is, in relation to the speed of crack growth (bond-snapping rate) [67]. According to Schmid's law [68, 69], cracks formed on {001} or {110} surfaces of a B1 crystal can be blunted via $\{110\}\langle1\bar{1}0\rangle$ and $\{111\}\langle1\bar{1}0\rangle$ slip inclined to the fracture plane. It is reasonable to assume that the rates of dislocation motion and crack growth are inversely related to the ideal shear and tensile strengths, respectively. Hence, the VN and VN$_{0.8}$ ability to withstand brittle cleavage on {001} and {110} planes can be assessed using the tensile-to-shear strength ratios m'·$\gamma_T^{\langle001\rangle}/\gamma_S^{\{110\}\langle1\bar{1}0\rangle}$ and m"·$\gamma_T^{\langle110\rangle}/\gamma_S^{\{111\}\langle1\bar{1}0\rangle}$, where m' = 0.5 and m" = 0.408 are the Schmid's factors for {110} and {111} slip planes with uniaxial loading along [001] and [110] directions, respectively. We suggest this ratio to be a realistic descriptor of crack resistance, analogously to the ratio of surface/unstable-stacking-fault formation energies used previously [70, 71].

The fact that the tensile/shear strength ratios calculated for VN$_{0.8}$ (≈0.7) are much larger than those (≈0.5) obtained for VN (see **Table 3**) is consistent with the superior fracture resistance of the understoichiometric compound (**Fig. 3**). We note, however, that our experimental results indicate that the toughness of VN$_{0.8}$ is *not* uniquely the result of plastic deformation. In fact, lattice slip in VN$_{0.8}$ occurs gradually, that is, to the extent necessary to prevent stress accumulation. This is demonstrated by absence of sudden pop-ins during experimental stress/strain measurements, **Fig. 2**, and by more moderate material pile-up subsequent to nanoindentation in comparison to VN, **Fig. 3c**. The fact that plastic flow (**Fig. 6**) and local structural transformations (**Figs. 4 and 5b**) in the understoichiometric compound become operative at elevated stress conditions is a requisite necessary for the simultaneous enhancement of hardness and toughness. The brittleness of VN is reflected by its lower tendency to activate lattice slip, as indicated by considerably larger shear strengths γs than in VN$_{0.8}$, and by the fact that stoichiometric phase undergoes sudden failure beyond its tensile yield points (**Fig. 5a**).

**3.3. Electronic mechanisms of enhanced hardness and toughness in VN$_{0.8}$.** DFT electronic-structure calculations provide fundamental insights for the electronic mechanisms which enhance hardness and toughness in VN$_{0.8}$. Energy-resolved electron densities, a technique for visualizing electronic states (see, e.g., Refs. [26, 72]), is here employed to demonstrate the enhancement of d-d metallic interactions in the vicinity of anion vacancies. **Fig. 7** illustrates the VN$_{0.8}$ electron densities resolved in energy intervals that primarily correspond to d-t$_{2g}$ (**Fig. 7a**) or d-e$_g$ states (**Fig.**

**7b**) near the Fermi level $E_F$. The energy ranges are identified by analyzing electronic densities of states (not shown). The figure demonstrates that constructive d-d wave interference (σ and π bonding states of $t_{2g}$ and $e_g$ symmetry, see bottom of **Fig. 7**) is favored by the presence of N vacancies ($N_{vac}$) both via $2_{nd}$ neighbor and $4_{th}$ neighbor (across $N_{vac}$) V–V orbital interactions. The observation that anion vacancies improve the metallic binding character in $VN_x$ has been previously suggested for lattices with isotropic vacancy ordering [73]. Due to high metallic nature, the electron density of $VN_{0.8}$ rearranges considerably when the material is subjected to external loads comparable to its ideal tensile $γ_T$ or shear $γ_S$ strengths. As detailed below, strain-mediated electron transfer both leads to modifications in bonding geometries during elongation and assists lattice slip upon shearing, thus enabling rapid stress dissipation.

**Fig. 8** illustrates the electron-transfer maps calculated for $VN_{0.8}$ under tensile deformation. In **Fig. 8a**, five adjacent atomic layers are labeled in alphabetic order from "a" to "e". V and N atomic sites along each plane are numbered sequentially from "1" to "4". Note that positions "2d" and "4d" are vacant. In unstrained $VN_{0.8}$(001), the metallic pairs $V_{2c}$–$V_{2e}$ and $V_{4c}$–$V_{4e}$ are linked by σ d-$e_g$ states across vacancy sites, as schematically represented on the bottom of **Fig. 7b**. For a uniaxial strain of 20%, the electrons that formerly occupied σ d-$e_g$ metallic states, transfer into nearest-neighbor $V_{2c}$–$N_{2b}$ and $V_{4c}$–$N_{4b}$ bonds parallel to the elongation, as well as $2_{nd}$-neighbor V–V σ d-$t_{2g}$ bonds orthogonal to the deformation direction. Overall, the substantial d-electron transfer promoted by the presence of vacancies in layer "d" produces (*i*) alternating weakened/reinforced V–N bonds at "a–b" and "b–c" layer interfaces and (*ii*) stronger metallic σ d-$t_{2g}$ binding within "a" and "c" V layers, which is evidenced by zig-zag patterned electron accumulation between all V–V pairs (**Fig. 8a**, 20% strain). **Fig. 8b** presents the $VN_{0.8}$ electronic-structure projected onto a {100} plane parallel to the elongation. The figure shows that the breakage of part of the bonds along the strain direction (note appearance of yellow-color regions in 20%-elongated $VN_{0.8}$(001), **Fig. 8b**) results in increased electron accumulation in the remaining (reinforced) vertical V–V and V–N bonds (note, e.g., $V_i$–$V_{ii}$–$N_i$–$V_{iii}$–$N_{ii}$ chain of bonded atoms in strained $VN_{0.8}$). The strain-mediated electronic mechanism leads to bond-bending and buckling of (001) layers, which confer on VN$_{0.8}$ a great fracture resistance.

Our stress/strain results indicate that the initially relatively higher resistance to shearing of VN$_{0.8}$, **Fig. 6**, originates from an overall stiffening of nearest-neighbor V–N bonds. The effect is due to a larger fraction of electrons that can be employed in d-e$_g$–p bonding states, analogous to the vacancy-induced strengthening demonstrated for B1 VMoN$_x$ alloys (see figure 7a in [7]). However, for δ$_{xz}$ > 0.09, the VN$_{0.8}$ response to shearing changes from hard to compliant, which facilitates lattice slip. A representative example is illustrated in **Fig. 9**, where electron transfer maps are calculated for a sequence of VN$_{0.8}$ atomic configurations sampled during {110}⟨1$\bar{1}$0⟩ slip. Time progression increases in alphabetic order from **Fig. 9a** to **Fig. 9i**. The (110) atomic layer labelled as "D" slides against the (110) layer "C". Lattice slip is assisted by continuous reorganization of d-electron clouds near to N$_{vac}$: back and forth transfer of electrons among σ d-e$_g$ V$_2$–V$_3$ fourth-neighbor bonds normal to the slip plane and σ d-t$_{2g}$ V$_1$–V$_3$ second-neighbor bonds parallel to the glide direction. In particular, a shortening of the σ d-e$_g$ V$_2$–V$_3$ bond lifts the V$_3$ atom upward (**Fig. 9c-g**), thus favoring lateral [$\bar{1}$10] translation of layer "D". Presumably, N diffusion also facilitates the crystal glide process, as suggested by the fact that, while the N$_1$ and N$_3$ atoms (see **Fig. 9a-d**) migrate out of the plane of view, two additional vacancies become visible upon completion of the slip process (compare **Fig. 9a** with **Fig. 9i**).

Electronic mechanisms (not shown), similar to those described in **Fig. 9**, are likely to facilitate {111}⟨1$\bar{1}$0⟩ slip in VN$_{0.8}$ (**Fig. 6b**). However, it may be expected that, during {111}⟨1$\bar{1}$0⟩ slip, B1 VN$_{0.8}$ domains locally characterized by N contents close to 0.5 are energetically inclined to form 111 stacking faults – with -A-B-C-B- sequence – as in hexagonal VN$_{0.5}$(0001) (space group P$\bar{3}$1m). This would provide additional degrees of freedom for VN$_{0.8}$ to dissipate mechanical stresses (analogous to the mechanism reported for metastable B1 Ti$_{0.5}$W$_{0.5}$N(111) solid solutions subjected to [1$\bar{1}$0] shearing [74]) as well as further enhance hardness by obstructing dislocation glide across the faults [75]. Last, it is worth noting that VN$_x$ (0.74<x<0.84) compounds present stable vacancy–ordered polymorphs [76]. Although lattice vacancies are randomly arranged in our films [19], $VN_{0.8}$ may locally activate disorder → order structural transitions in response to external loading, leading to an alternative form of transformation toughening.

## 4. Summary and perspectives

We combine experiments and first-principles simulations to demonstrate that the control of anion vacancy concentrations in epitaxial single-crystal B1 $VN_x(001)/MgO(001)$ ceramics allows simultaneous enhancement of both material's hardness and toughness.

Berkovich nanoindentation, used to measure the films' hardness, shows that understoichiometric $VN_{0.8}$ is ≈20% harder than stoichiometric VN samples. The materials' fracture toughness is qualitatively assessed by cube-corner nanoindentation, performed at constant penetrations which largely exceed the films' thickness. These tests demonstrate that, while VN fractures in a brittle manner, the understoichiometric $VN_{0.8}$ compound *never* cracks.

First-principles atomistic modeling of supercells subjected to tensile and shear deformation allows us rationalizing the dramatic differences in the mechanical behavior of VN vs. $VN_{0.8}$. Our results show that $VN_{0.8}$ possesses higher elastic shear stiffness than VN, which may explain its greater measured hardness. However, despite its initially higher resistance to change of shape, $VN_{0.8}$ requires considerably lower shear stresses than VN to induce $\{110\}\langle1\bar{1}0\rangle$ and $\{111\}\langle1\bar{1}0\rangle$ lattice slip at room temperature. For supercells subjected to uniaxial elongation, AIMD simulations show that, while stoichiometric VN crystals suddenly cleave at their yield point, the understoichiometric compound activates local lattice transformations, which reflect its superior toughness.

Thorough analyses of the $VN_{0.8}$ electronic structures indicate that (*i*) the high elastic resistance to shearing originates from intrinsically stronger first-neighbor V–N bonds, whereas (*ii*) the ability to dissipate accumulated external stresses by inducing structural transformations and lattice slip stems from the possibility of mutually transferring electrons among $2_{nd}$- and $4_{th}$-neighbor (across vacancy) V–V metallic states. In contrast, the relative softness of the stoichiometric compound originates from electronic/phonon instabilities which facilitate B1→tetragonal martensitic phase transitions upon shearing in the elastic-response regime.

It would be reasonable to assume that a vacancy-induced enhancement in metallic character of anion-deficient B1 TM carbonitrides – evidenced by the increased x-ray photoelectron spectra intensities at low binding energies [9, 77-81] – necessarily implies improved plasticity. However, rigorous experimental testing should be performed to verify whether the material's toughness is effectively increased by an enhanced metallic-bonding character, i.e., exclude a concomitant drop in hardness. In this regard, although experimental results consistently show that anion vacancies harden B1 Group-VB carbonitrides, conflicting trends in H vs. vacancy concentration have been reported for Group-IVB carbonitrides [16, 28, 82-84]. Rationale design of hard and tough carbonitrides should aim at triggering the material's plastic response at the *right stage* of a deformation process: *too early* would imply softness; *too late* may cause brittle fracture. We foresee that the concentration of valence electrons is a key parameter to control the hard → plastic turning point under loading. This study offers a novel strategy to identify alloy and/or multicomponent (high-entropy) refractory carbides and nitrides with superior combination of hardness and toughness.

**Acknowledgements**
All simulations were carried out using the resources provided by the Swedish National Infrastructure for Computing (SNIC) – partially funded by the Swedish Research Council through grant agreement no. 2016-07213 – on the Clusters located at the National Supercomputer Centre (NSC) in Linköping, the Center for High Performance Computing (PDC) in Stockholm, and at the High Performance Computing Center North (HPC2N) in Umeå, Sweden. We thank N. Koutná (TU Wien) for useful discussions. D.G.S. gratefully acknowledges financial support from the VINN Excellence Center Functional Nanoscale Materials (FunMat-2) Grant 2016–05156 and from the Olle Engkvist Foundation. L.H. acknowledges the Knut and Alice Wallenberg Foundation for a Scholar Grant (KAW-2016-0358) and, together with I.P. and J.E.G., also the Swedish Government Strategic Research Area in Materials Science on Advanced Functional Materials at Linköping University (Faculty Grant SFO-Mat-LiU No. 2009-00971).

deformation ($\delta_{ij}$) to a supercell with generic orientation, are determined by appropriate rotations (R) of σ and δ matrixes. The stress (σ') and strain (δ') tensors in a coordinate system with ⟨001⟩ crystallographic axes parallel to x, y, and z Cartesian directions are calculated via: σ' = R·σ·R$_{-1}$ and δ' = R·δ·R$_{-1}$. Note that two different rotations need to be applied in the case of deformation applied to supercells with [111] vertical orientation.

**Tables and figures**

|  | $\gamma_{T/S}$ (GPa) | $\delta_{T/S}$ (%) | $U_T$ (GPa) |
|---|---|---|---|
| **VN** | | | |
| Tensile ⟨001⟩ | 36 | 14 | 3.6 |
| Tensile ⟨110⟩ | 46 | 16 | 4.7 |
| Shear {110}⟨1$\bar{1}$0⟩ | 34 | 18 | |
| Shear {111}⟨1$\bar{1}$0⟩ | 39 | 22 | |
| **VN$_{0.8}$** | | | |
| Tensile ⟨001⟩ | 32 | 30 | 7.6 |
| Tensile ⟨110⟩ | 42 | 23 | 6.2 |
| Shear {110}⟨1$\bar{1}$0⟩ | 23 | 14 | |
| Shear {111}⟨1$\bar{1}$0⟩ | 25 | 18 | |

**Table 1.** Ideal tensile and shear strengths $\gamma_T$ and $\gamma_S$, tensile toughness $U_T$, and yield strains $\delta_T$ and $\delta_S$ of B1 VN$_x$ crystals determined by AIMD simulations at 300 K.

|  | $C_{11}$ (GPa) | $C_{12}$ (GPa) | $C_{44}$ (GPa) | $G_{110}$ (GPa) | $G_{111u}$ (GPa) |
|---|---|---|---|---|---|
| **VN** | | | | | |
| **Actual** stiffness $C_{ij}$, $G_{hkl}$ | **606±8** | - | * | **116±13** | **94±18** |
| *Effective* stiffness $\widetilde{C}_{ij}$, $\widetilde{G}_{hkl}$ | 603±20 | 258±37 | 135±18 | 172±21 | 193±15 |
| Exper. & DFT | 533–623 | 135–251 | 122–196 | 186–325 | 158–266 |
| **VN$_{0.8}$** | | | | | |
| **Actual** stiffness $C_{ij}$, $G_{hkl}$ | **476±15** | - | * | **205±7** | **190±7** |
| *Effective* stiffness $\widetilde{C}_{ij}$, $\widetilde{G}_{hkl}$ | 509±36 | 168±22 | 111±24 | 171±21 | 151±16 |
| Exper. & DFT | 512–555 | 113–165 | 127–240 | 195–315 | 164–285 |

**Table 2.** Elastic constants $C_{ij}$ and shear moduli $G_{hkl}$ of B1 VN and VN$_{0.8}$ determined by present AIMD simulations at room temperature [57] vs. previous experimental and 0-K DFT results [18, 85-94]. The statistical uncertainty on $C_{11}$, $G_{110}$, and $G_{111u}$ values accounts for stress fluctuations during AIMD. *Not calculated: direct evaluation of the *actual* $C_{44}$ [$\equiv G_{001}$] shear stiffness requires modelling {001}⟨1$\bar{1}$0⟩ shear deformation.

|  | m'·$\gamma_T^{⟨001⟩}/\gamma_S^{\{110\}⟨1\bar{1}0⟩}$ | m''·$\gamma_T^{⟨110⟩}/\gamma_S^{\{111\}⟨1\bar{1}0⟩}$ |
|---|---|---|
| VN | 0.53 | 0.48 |
| VN$_{0.8}$ | 0.70 | 0.69 |

**Table 3.** Calculated tensile-to-shear strength ratios, here used as indicators of fracture resistance.

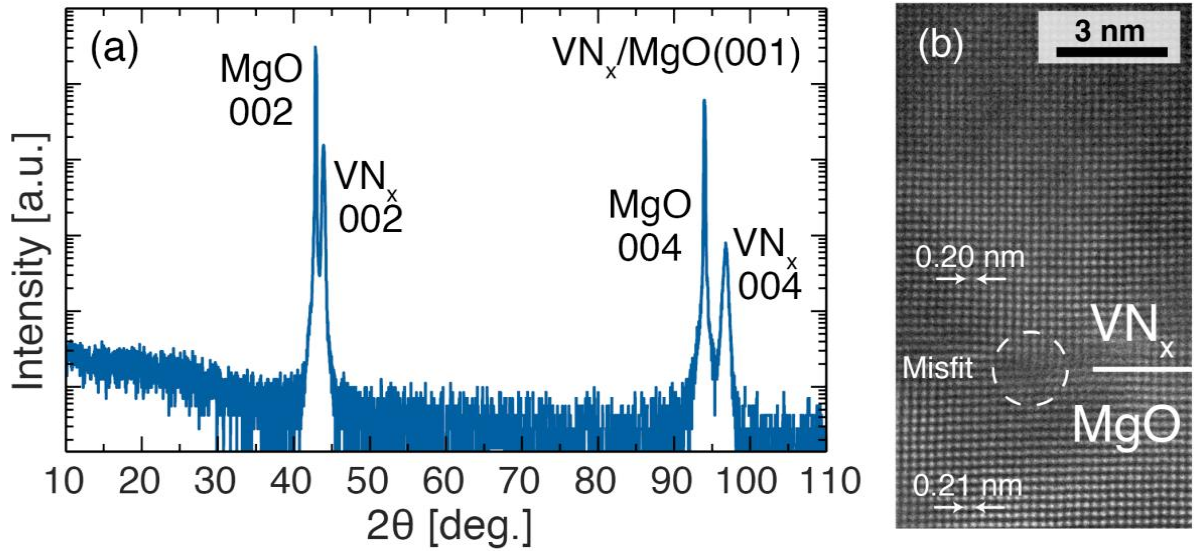

**Fig. 1. (a)** XRD ω-2θ scan from a 300-nm-thick understoichiometric VN$_{0.8}$/MgO(001) layer. Four peaks are present, consistent with an epitaxial single-crystalline film. **(b)** HR-XTEM micrograph acquired near the film/substrate interface and along the [100] zone axis of a VN/MgO(001) film. A misfit dislocation is encircled in panel **(b)**.

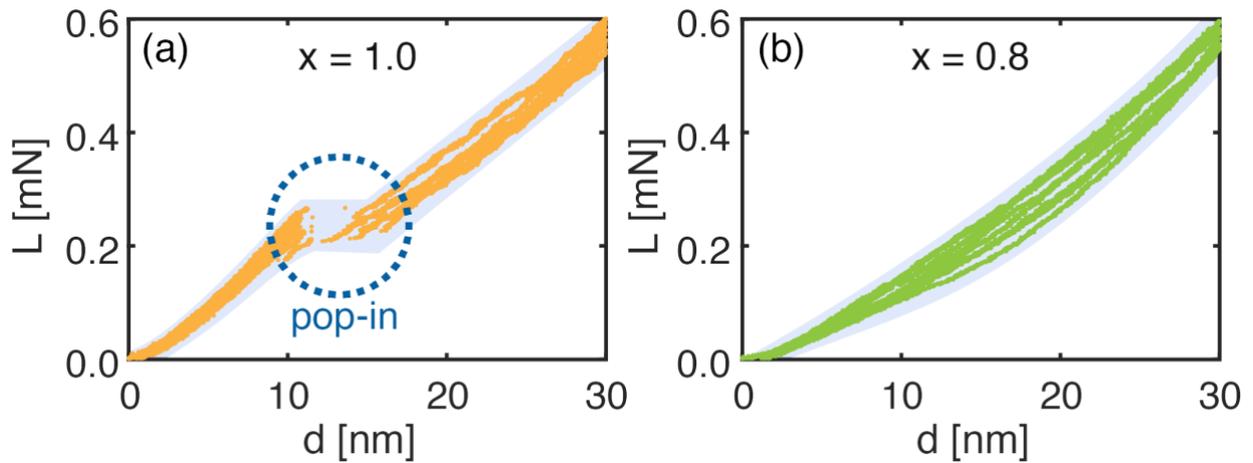

**Fig. 2.** Typical load-displacement curves L(d) obtained from nanoindentations of 300-nm-thick VN$_x$/MgO(001) films with **(a)** x = 1.0 and **(b)** x = 0.8. Pop-ins signaling catastrophic material failure at the mesoscopic scale are observed only in stoichiometric films.

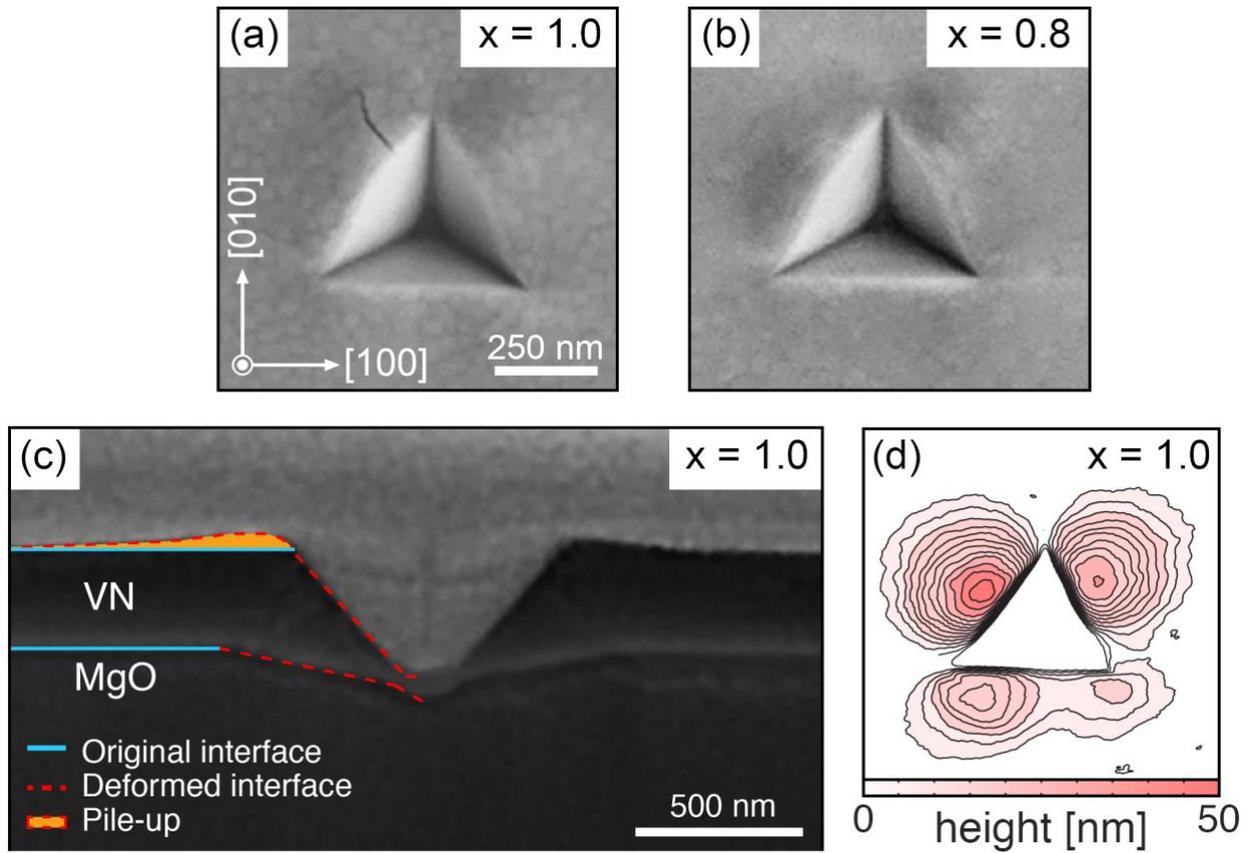

**Fig. 3.** SEM images of 400-nm-deep cube-corner indentations of 300-nm-thick $VN_x$/MgO(001) layers with N contents **(a)** x = 1.0 and **(b)** x = 0.8. **(c)** Cross-sectional SEM image of the indent in panel **(a)**. The vertical section is cut using a focused $Ga_+$ ion beam. **(d)** SPM height isointensity contours of the indent in panel **(a)** showing material pile-up along <110> directions. The mounds retain their X-pattern while the triangular indentor probe is rotated around the indentation axis. The results shown for VN/MgO(001) in **(d)** are also representative for $VN_{0.8}$/MgO(001).

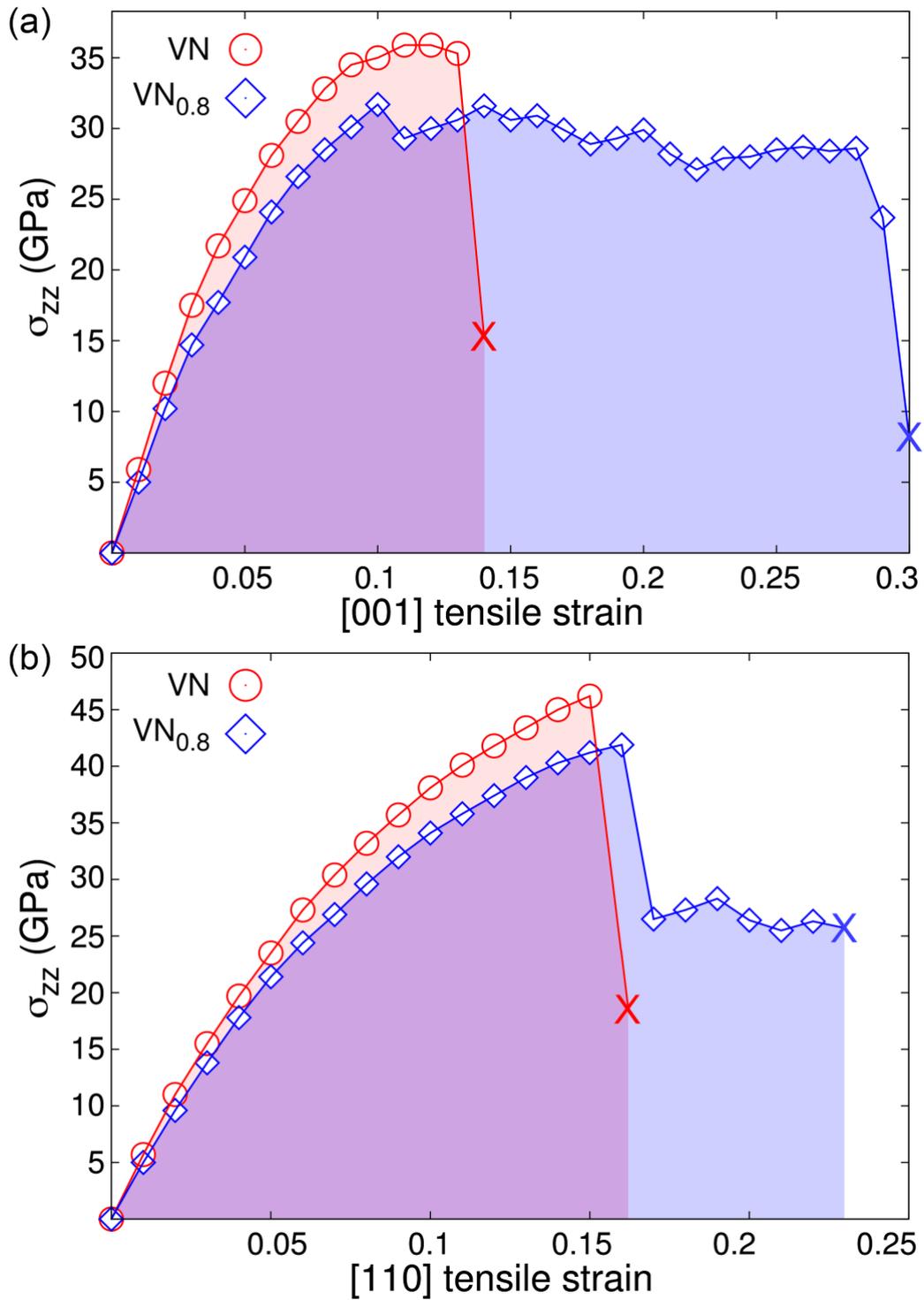

**Fig. 4.** Tensile stress $\sigma_{zz}$ determined as a function of tensile strain along **(a)** [001] and **(b)** [110] directions by AIMD simulations at 300 K for VN and VN$_{0.8}$. Fracture points are indicated by "×" symbols. For both elongation directions, the understoichiometric compound exhibits superior toughness (indicated by colored shaded areas) than VN.

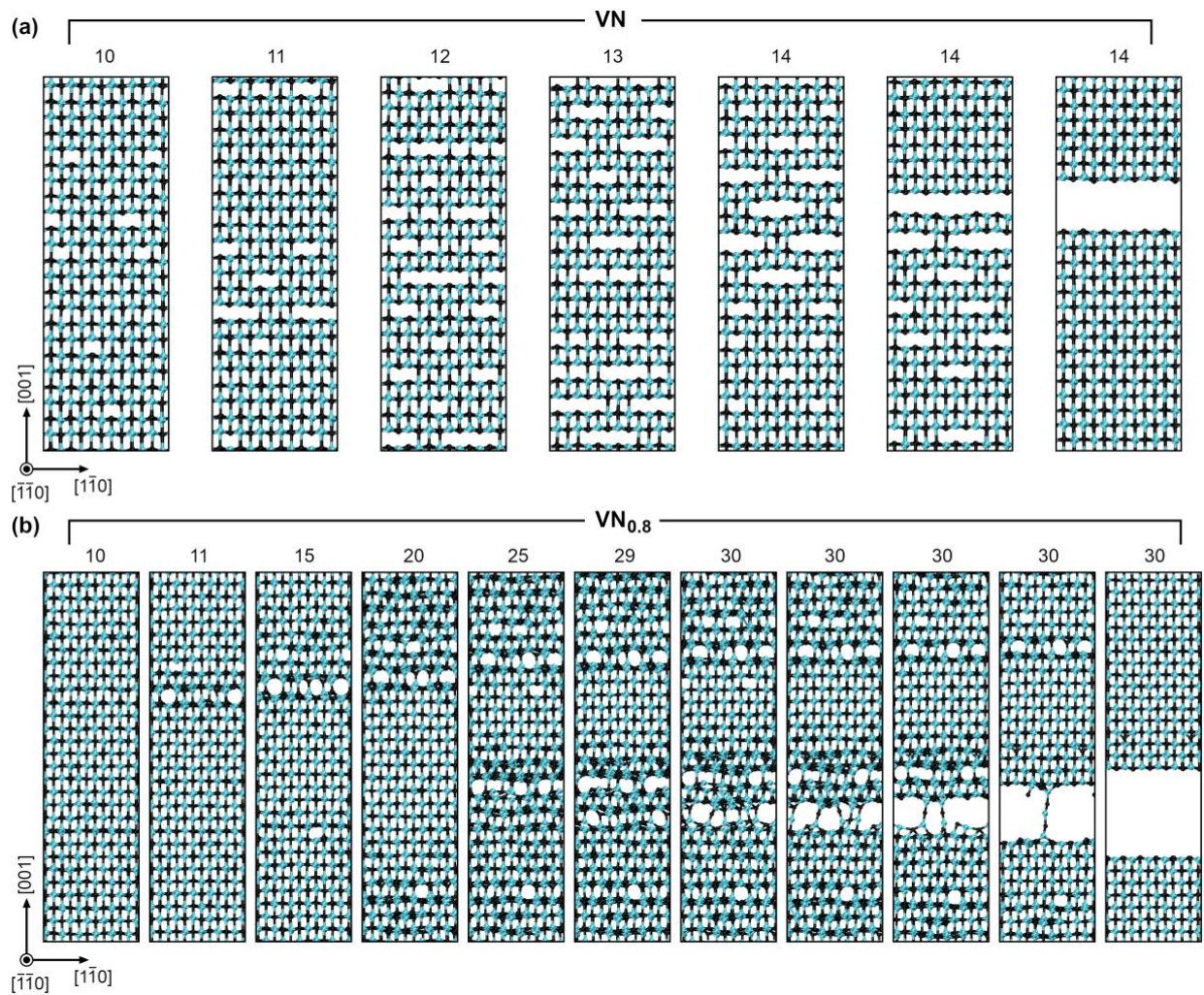

**Fig. 5.** AIMD snapshots during [001] tensile-loading of **(a)** stoichiometric VN(001) and **(b)** understoichiometric VN$_{0.8}$(001). Vanadium and nitrogen atoms are represented as cyan and black spheres. Each panel is labeled by numbers that indicate elongation percentages δ$_{zz}$. Cleavage of the crystals on (001) planes can be visualized in a sequence of snapshots at constant strains, separated by fractions of ps. While VN fractures at δ$_T$ = 14%, understoichiometric VN$_{0.8}$ sustains elongation up to 30% without cracking. Strain-mediating structural transformations in VN$_{0.8}$ result in enhanced (dislocation-free) plasticity.

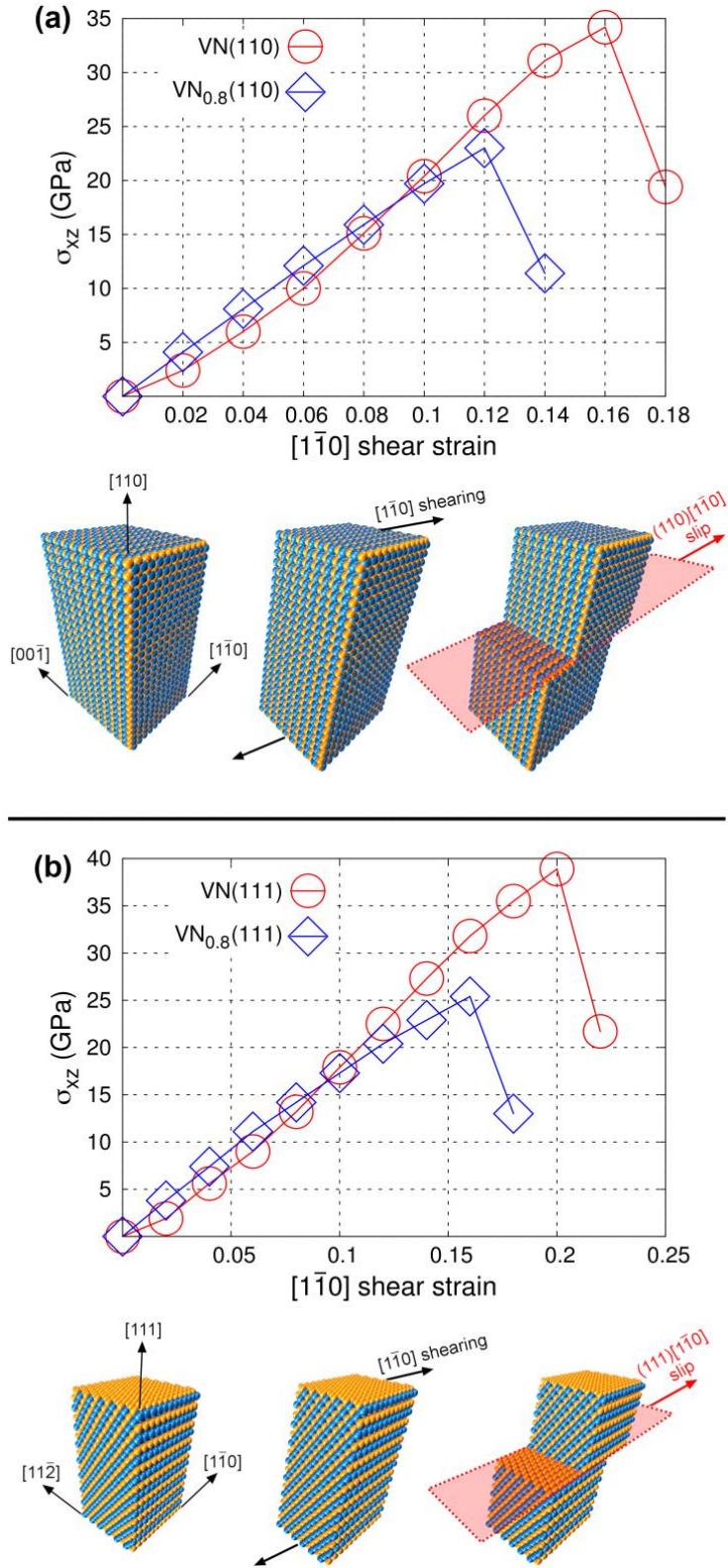

**Fig. 6.** AIMD stress/strain curves calculated at 300 K during shearing of **(a)** VN(110) and VN$_{0.8}$(110) and **(b)** VN(111) and VN$_{0.8}$(111) along the Burgers vector direction [1$\bar{1}$0] up to the occurrence of lattice slip (note drops in shear stress σ$_{xz}$ at curve extremities). {110}⟨1$\bar{1}$0⟩ and {111}⟨1$\bar{1}$0⟩ shearing and lattice slip are schematically illustrated in the lower part of each panel. For both {110}⟨1$\bar{1}$0⟩ and {111}⟨1$\bar{1}$0⟩ slip systems, VN$_{0.8}$ exhibits an initially stronger resistance to shear deformation (up to ≈9%), followed by more facile activation (lower ideal shear strengths) of layer glide than calculated for the stoichiometric compound.

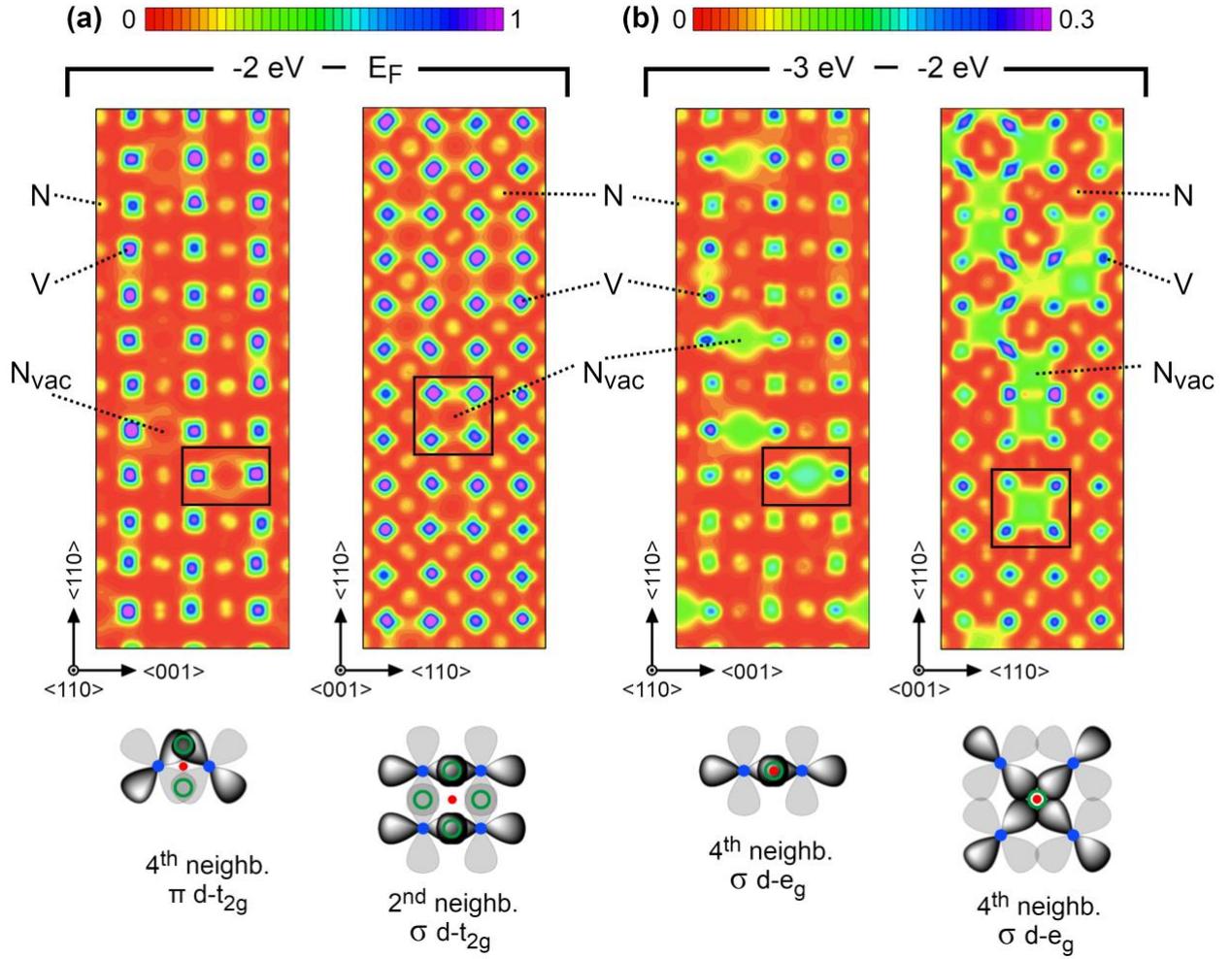

**Fig. 7.** Energy-resolved electron density maps of unstrained VN$_{0.8}$. The figure panels illustrate electronic states which primarily correspond to **(a)** d-t$_{2g}$ [–2 eV - E$_F$] and **(b)** d-e$_g$ [–3 - –2 eV] metallic bonds near the Fermi level E$_F$. Both panel **(a)** and **(b)** provide examples of electron density distributions on {110} and {001} crystallographic planes. The black squared frames facilitate visualization of different chemical bonds, which are also schematically represented by d-d orbital overlapping (see lower part of the figure). In illustrations at the bottom, blue and red dots mark V nuclei and anion-vacancy (N$_{vac}$) positions, respectively, whereas green circles indicate centers for constructive d-d electronic-wave interference. Note that, while **(a)** π and σ d-t$_{2g}$ metallic bonding accumulate charge around the vacancy site, **(b)** σ d-e$_g$ V – V bonding across vacancy sites yields electron accumulation on N$_{vac}$. The color scales are expressed in e$^-$·Å$^{-3}$.

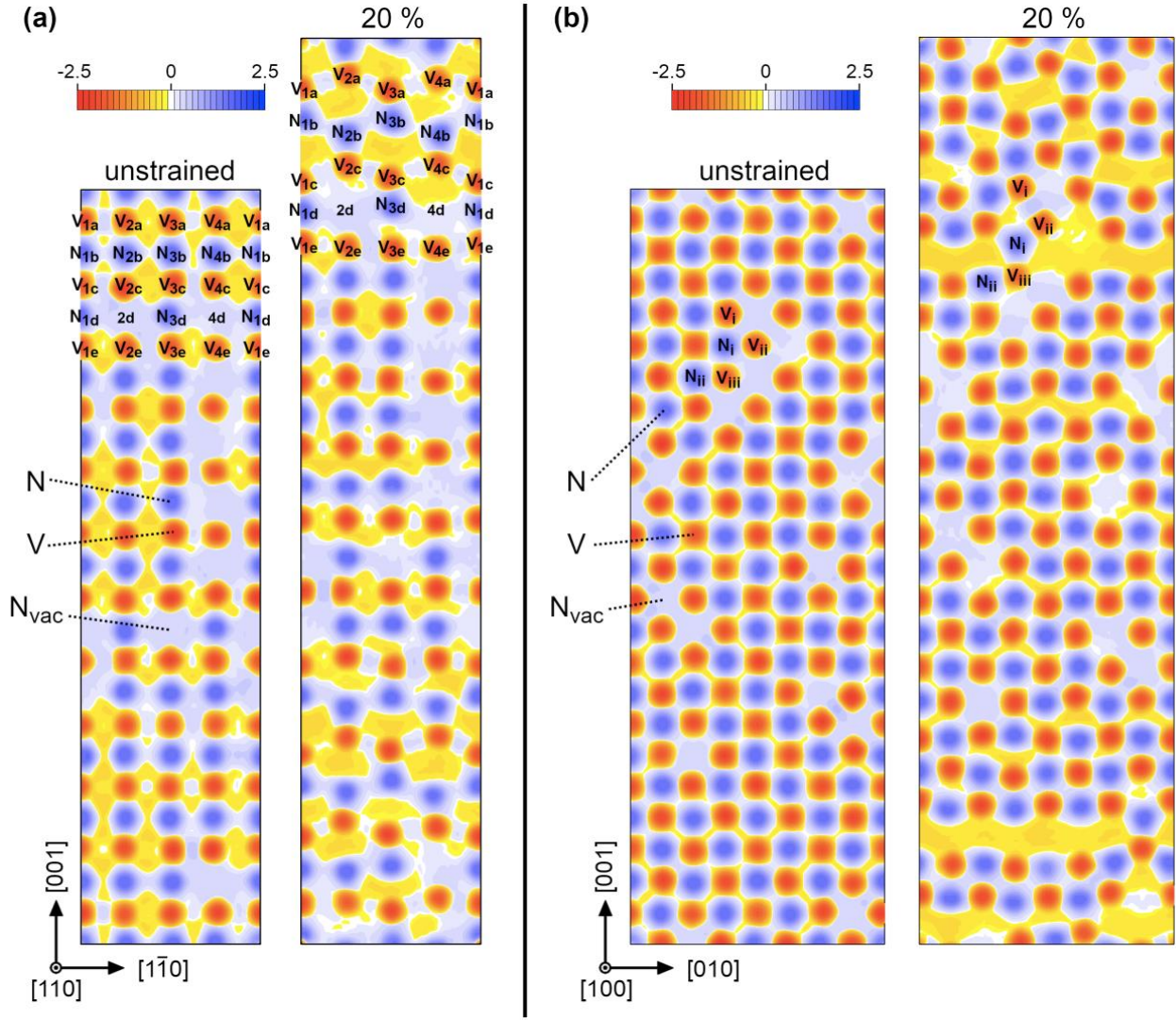

**Fig. 8.** Electron transfer maps calculated for unstrained and 20% tensile-strained $VN_{0.8}(001)$ at 300 K. Examples of electron distributions are shown for **(a)** {110} and **(b)** {100} crystallographic planes parallel to the elongation direction. Electron accumulation (depletion) is indicated in blue (red-yellow) colors, with scale expressed in $e^- \cdot \text{Å}^{-3}$.

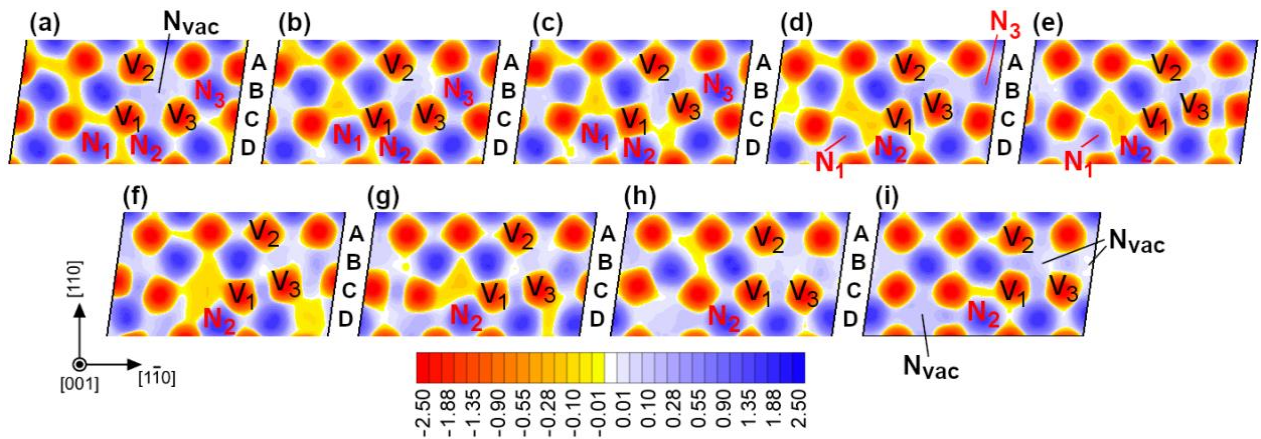

**Fig. 9.** Electron transfer maps calculated for a sequence of $VN_{0.8}(110)$ configurations during $\{110\}\langle 1\bar{1}0\rangle$ lattice slip (i.e., constant shear strain $\delta_{xz} = 14\%$, see **Fig. 6a**) at 300 K. Note that the figure shows only atomic layers near the glide-plane. The color scale is expressed in $e^- \cdot \text{Å}^{-3}$.

**Graphical abstract**

Increased hardness and toughness
in single-crystal B1 VN$_{0.8}$

Experimental evidence

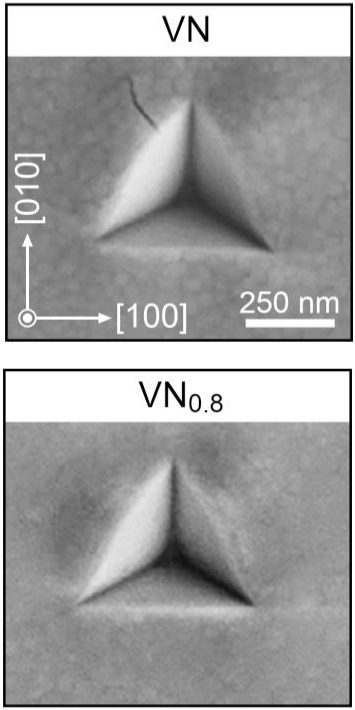

Electronic-level understanding

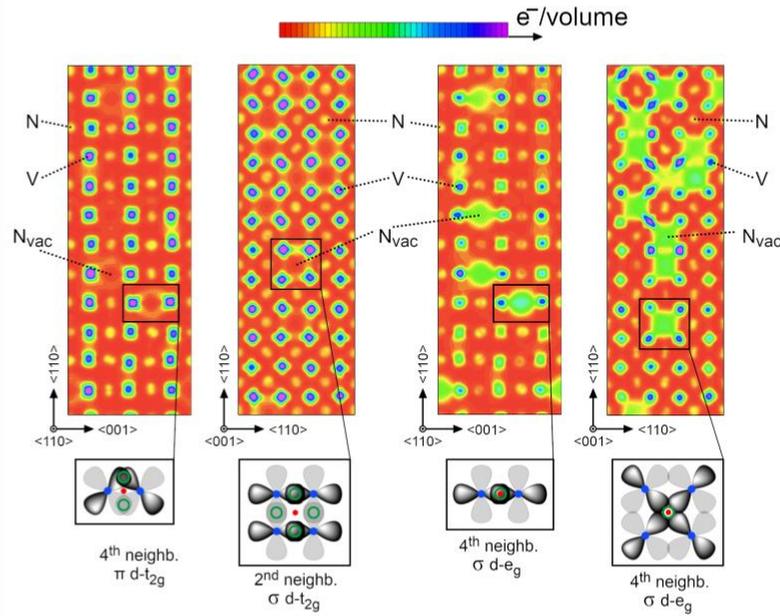

Atomistic pathways and mechanisms

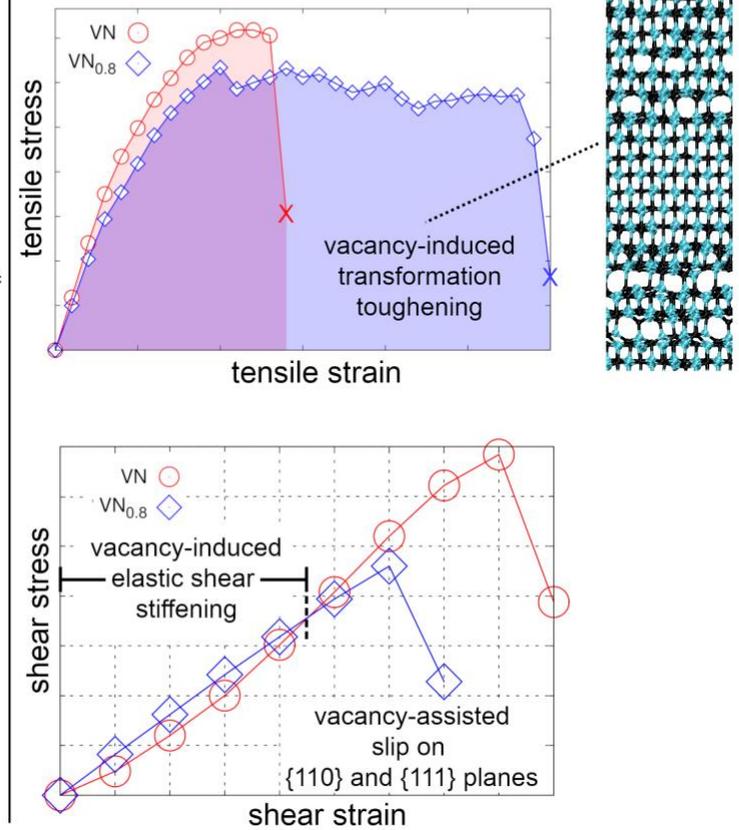